# Spatially non-homogeneous metallization of VO$_2$:

# a TDDFT+DMFT analysis

Jose Mario Galicia- Hernandez,[a,b] Volodymyr Turkowski,[a] Gregorio Hernandez-Cocoletzi[b] and Talat S. Rahman[a,*]

[a]Department of Physics, University of Central Florida, Orlando, FL 32816, USA
[b]Instituto de Fisica Ing. Luis Rivera Terrazas, Benemerita Universidad Autonoma de Puebla, Puebla 72550, Mexico,
*Corresponding Author, e-mail address: Talat.Rahman@ucf.edu

ABSTRACT. We provide insights into the atomistic details of the ultrafast spatially-resolved breakdown of the insulating M$_1$ phase in bulk VO$_2$ employing an *ab initio* technique based on time-dependent density-functional theory and dynamical mean-field theory (DMFT-TDDFT). We find that the system is initially metallized preferentially along the vanadium-dimer chains (C$_R$ axis), with a subsequent growth of C$_R$-elongated metallic bubbles. Moreover, we trace the breakdown of the insulating phase to two types of oxygen atoms, resulting from vanadium dimerization, which produce an unusual charge-density modulation in the oxygen-atom chains with significant charge transfer to the inter-dimer distance. These results are in qualitative agreement with experimental data and shed light on the interplay between valence charge and lattice structure and its role in the ultrafast response of strongly correlated insulators.

*Introduction.* – After the fundamental work of Cavalleri and collaborators,[1] who demonstrated that the insulator-to-metal transition (IMT) in VO$_2$ excited by an ultrafast laser pulse happens at the fs time scale, it became clear that the electronic properties of strongly correlated materials can be tuned at the ultrafast time scale. In addition to new interesting physical phenomena accompanying the ultrafast IMT in these materials, one may also expect many practical applications of the effect, from switches,[6-8] to biosensors.[11] Thus, microscopic understanding of the ultrafast properties of VO$_2$, one of the prototype strongly correlated materials, is of fundamental and technological importance. Not surprisingly, a large amount of the experimental data on the IMT in VO$_2$ has been accumulated (see Ref. [22] and references therein). In particular, it is well-established that at laser-pulse fluences below 8 mJ/cm$^2$ the IMT in the system happens without the lattice transformation, i.e. the insulating phase monoclinic (M$_1$) crystal structure does not change, while the system becomes conducting due to transfer of electrons to the conducting bands. Such a new metallic phase, called $\mathcal{M}$ phase, is characterized also by collapse of the bandgap and rather long lifetimes (100-300fs) of excited electrons and holes [2]. There are also serious indications that the IMT in VO$_2$ may have **spatially-nonhomogeneous** character. In particular, signatures of the coexistence of metallic and insulating domains during the transition have been reported [22,24,25,26]. For example, it has been shown [25] that the system illumination may lead to a metallization along narrow filaments. Similarly, Hilton et al. [26] suggested the existence of metallic domains created by an ultrafast pulse excitations in their optical-pump terahertz-probe spectroscopy studies. They also estimated the time-dependent size of these domains from their results for the nonequilibrium conductivity. Similarly, in the spatially-resolved low-fluence pulse-induced metallization ($\mathcal{M}$ phase) studies of $VO_2$ with a combined ultrafast electron diffraction and time-resolved terahertz spectroscopy approach, it was shown [22] that in the $\mathcal{M}$ phase there is a periodic (anti-ferroelectric) modulation in the electrostatic crystal potential on the oxygen sites along the ("V-dimer") $C_R$ axis, commensurate with the lattice constant. The largest change in the potential takes place on the oxygen atoms that are related to the minimum V–O distance in the octahedra (and to the V–V dimer tilt). Moreover, significant changes in the electrostatic potential were also observed between the vanadium atoms in the octahedrally-coordinated chains along $C_R$, indicating a transfer of charge from the vanadium dimers to the inter-dimer region. Recent few-femtosecond extreme UV transient absorption spectroscopy (FXTAS) data suggest that the IMT in $VO_2$ may take place in much shorted, tens of fs, times and moreover, often neglected O-p states may play an active role in the charge dynamics. Our targets in this study is to identify microscopic spatially-resolved femtosecond charge dynamics and how it affects the metallization of the system.

To identify the microscopic factors that control the response of the system with strong electron-electron correlations is a great challenge for the theory. For the first time, a systematic study of the breakdown of the insulating phase in $VO_2$ was reported by He and Millis[30] who solved the problem in two stages: the static properties of the system were analyzed by using DFT+U and the excited properties of the system were studied by using a Boltzmann equation approach. Though the reported results contain several important results, like established time scales for different stages of the system response and existence of a metastable metallic $\mathcal{M}$ phase (with the insulating ($M_1$) lattice structure), there remain some questions on the accuracy of the used approximation – momentum-averaged Coulomb interaction and neglect of possible effects of memory and of spatially-nonhomogeneous charge dynamics.

All these effects can be easily included into the analysis by using a much less computationally expensive TDDFT approach, provided one uses an appropriate XC potential to properly include the electron-electron correlations. Recently, we have developed an algorithm how to derive the linear-response XC potential (i.e., XC kernel) from dynamical mean-field theory (DMFT).[33] Namely, the XC kernel is calculated from the charge susceptibility for the effective Hubbard model for given material. Beside the low computational cost and inclusion of the memory and spatial-fluctuation (nonhomogeneity) effects, the approach has another advantage as compared to the DFT+U algorithm: it takes into account the time-resolved on-site charge interactions already at the static level (through the static DMFT calculations). As result, contrary to DFT+U, DMFT gives *nonmagnetic* insulating ground state in the $M_1$ phase (see, e.g., Ref.[18] for the DFT+U results) and *metallic* electronic structure in the rutile phase (i.e., DMFT is capable to describe the IMT with temperature induced lattice transformation, while DFT+U gives insulating phase for both M1 and rutile lattice structures).

In this work, we perform analysis of the ten(s)-fs spatially-resolved charge dynamics in $VO_2$ with TDDFT+DMFT in order to answer the following questions:

1) The origin for the anti-ferroelectric charge order on the oxygen atoms and for the spatially-nonhomogeneous modifications of the electrostatic potential on the vanadium atoms after the excitation.

2) The time-dependence of the metallic domains and its dependence on the memory effects and strength of the on-site Coulomb interaction and how these effects change the ultrafast conductivity.

*TDDFT+DMFT approach and the computational details.* – We began with the DFT calculations by using the Quantum Espresso package[35] with the Perdew-Wang LDA XC potential, norm-conserving semi-local pseudopotentials, $9 \times 9 \times 9$ Monkhorst–Pack k-points mesh in the first Brillouin zone and cut-off energy 70 Ry to get the initial static results: Kohn-Sham electron eigen-energies and eigen-functions and the relaxed lattice structure. The correlation effects corrections were included into by performing the DMFT calculations with the perturbation theory (MO-IPT) impurity solver[34] (in the framework of so called DFT+DMFT scheme[34]; the used interaction parameters: U=4eV, J=0.65eV, intra-dimer repulsion V=1eV). As the last step, the time-dependent analysis was performed with TDDFT with the DMFT XC kernel. Details of derivation of the XC kernel are given in Supplementary Information (SI), Section A. The density-matrix TDDFT formalism to calculate the excited charge dynamics is provided in SI, Section B and details of our approach to study of the spatially-resolved domain growth - in SI, Section C. We present also details of used Bruggeman's formalism to calculate the conductivity as function of the excited metallic volume in SI, Section D. Finally, details of the calculations of the spatially-resolved excited charge density and of the resulting change of the electrostatic potential are given in SI, Section E (for more details of the calculations, see the Supplementary Information and our recent paper [42]).

*Ground-state properties.* – The results for the projected (most optically-active) vanadium-atom d-orbital DFT DOS are shown in Fig.1b.

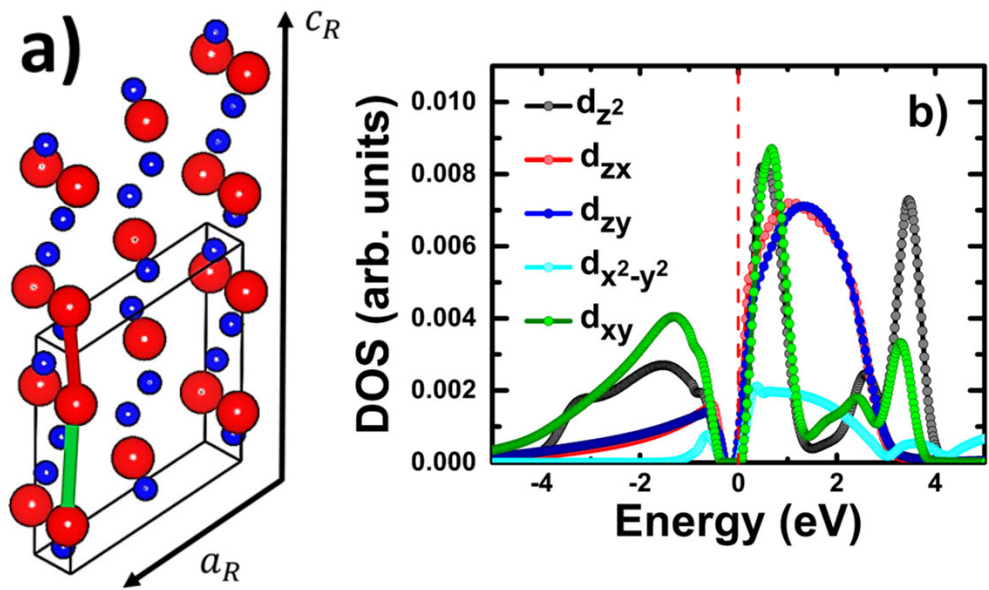

**Figure 1.** (a) The $M_1$ unit cell with the vanadium and oxygen atoms shown in red and blue, correspondingly. A pair of V-dimer atoms along the $C_R$ axis is marked by a green bond, while two unpaired atoms – by the red bond. (b) V-atom d-orbital DOS obtained with the DFT+DMFT calculations.

It is important to stress that to open the experimentally-observed (0.67eV) gap in the DOS (Fig.1) one needs to take into account two effects: on-site and the inter-site within the vanadium dimers[14] Coulomb interaction on the vanadium d-orbitals (the dimerized lattice structure of $VO_2$ in the $M_1$ phase is shown in Fig.1a).

*"Long-scale" (nm) nonhomogeneous response.* - We performed an analysis of the time-dependencies of the metallic domain radius at different values of the Coulomb repulsion (Figs.2a, b). These were calculated by finding the space and time dependencies of the correlation function for the Kohn-Sham wave functions from the TDDFT equation (for details of the scaling approach, see SI, Section C and, e.g., Ref. [36]). As results of our calculations show, the increase of the Coulomb interaction U "accelerates" the initial growth due to enhanced scattering effects. Also, due to different effective hopping parameters along and perpendicular to the V chains, the metallic bubbles are elongated in the chain direction (Fig.2d). Again, the nonadiabaticity effects play an important role in the domain growth (compare the blue ("nonadiabatic") curve in Figs. 2a and ("adiabatic") curves In Fig. 2b for small value of U (we did not show results for larger U's in the adiabatic case, where the bubble radius is growing even much faster than in the nonadiabatic case)). The power-like growth of the metallic domains in time results in similar anisotropic growth of conductivity that grows as cube of the domain size (Fig. 2c). The results for conductivity were obtained by using Brugemann's theory (SI, Section D). Also, at large U's (blue curve in Fig.2a), one observes oscillatory features in conductivity at short times that are consequence of the oscillatory time-dependence of the XC kernel (i.e., memory effects play an important role).

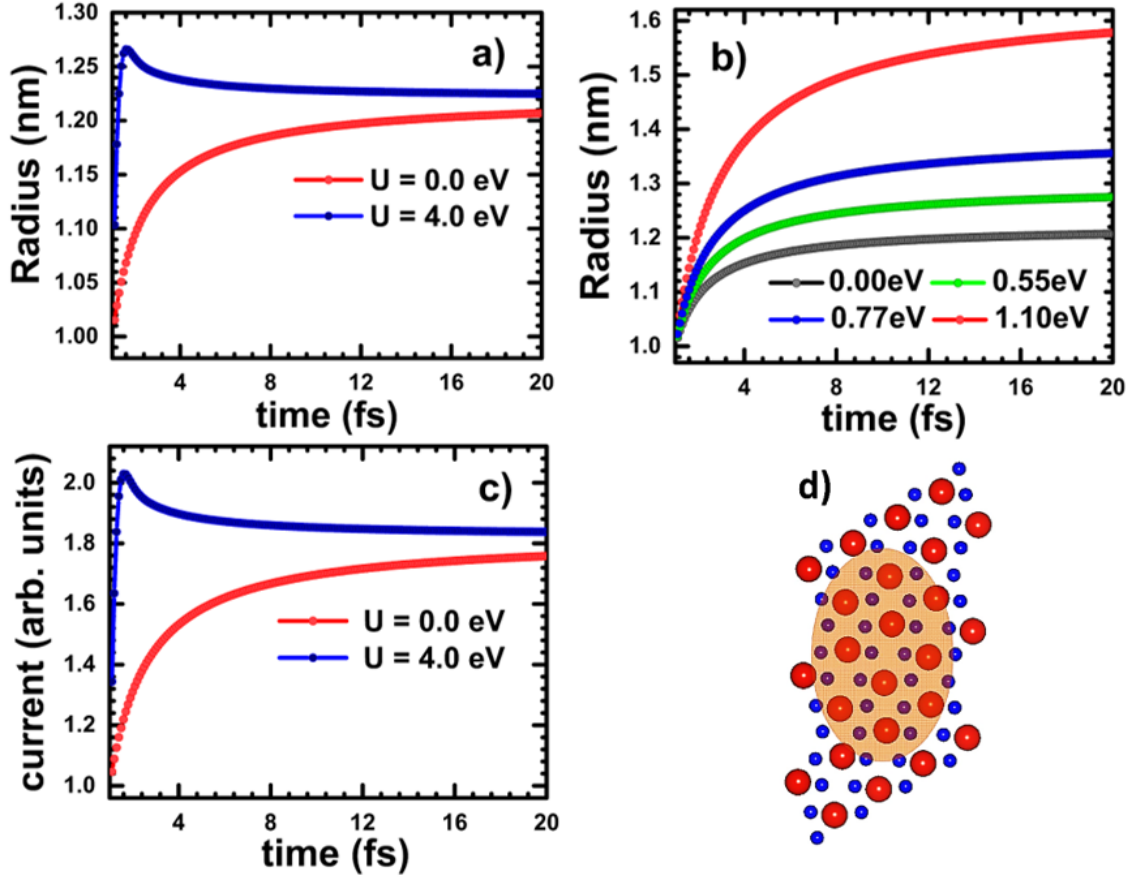

**Figure 2.** The metallic domain radius (in the chain direction) as function of time for the non-adiabatic (a) and adiabatic (b) approximations and the corresponding to (a) fs conductivity (proportional to the domain volume) (c) in the case of different values of the Coulomb repulsion parameter, field magnitude $E = 2\frac{eV}{nm}$ and pulse duration 1fs, that corresponds to fluence 0.4mJ/cm$^2$. d) graphic representation of the shape of a metallic bubble elongated by ~50% in the direction of vanadium chains.

As it follows from Fig.2, at initial times the conductivity is proportional to the cube of the metallic domain radius, or $t^{3/2}$. Such a dependence is in a qualitative agreement with the short-time part of the available experimental data (though the time-resolution of the data in these experiments is too large, ~10fs; see, for example, Ref. [26]). To test possible oscillations in conductivity experimentally, shorter resolution times and possibly rather low temperatures need to be used.

*Sub-nm spatially-resolved dynamics.* - To understand the spatially-resolved scenario of the of metallization in VO$_2$ at shorter, sub-nm, scale we have taken a "closer" look at the projected DOS for different atoms. As our results show, there are two nonequivalent oxygen atoms in the unit cell. Namely, there are atoms close to the V dimer and close to the inter-dimer spacing (blue and green atoms in Fig.3a, correspondingly). This non-equivalence is reflected in different DOS for non-equivalent oxygen atoms (Figs. 3b and 3c). Namely, there are more available states on the "dimer-close" oxygen atoms, which results in the anti-ferroelectric ordering of the oxygen subsystem in the excited state.[22]

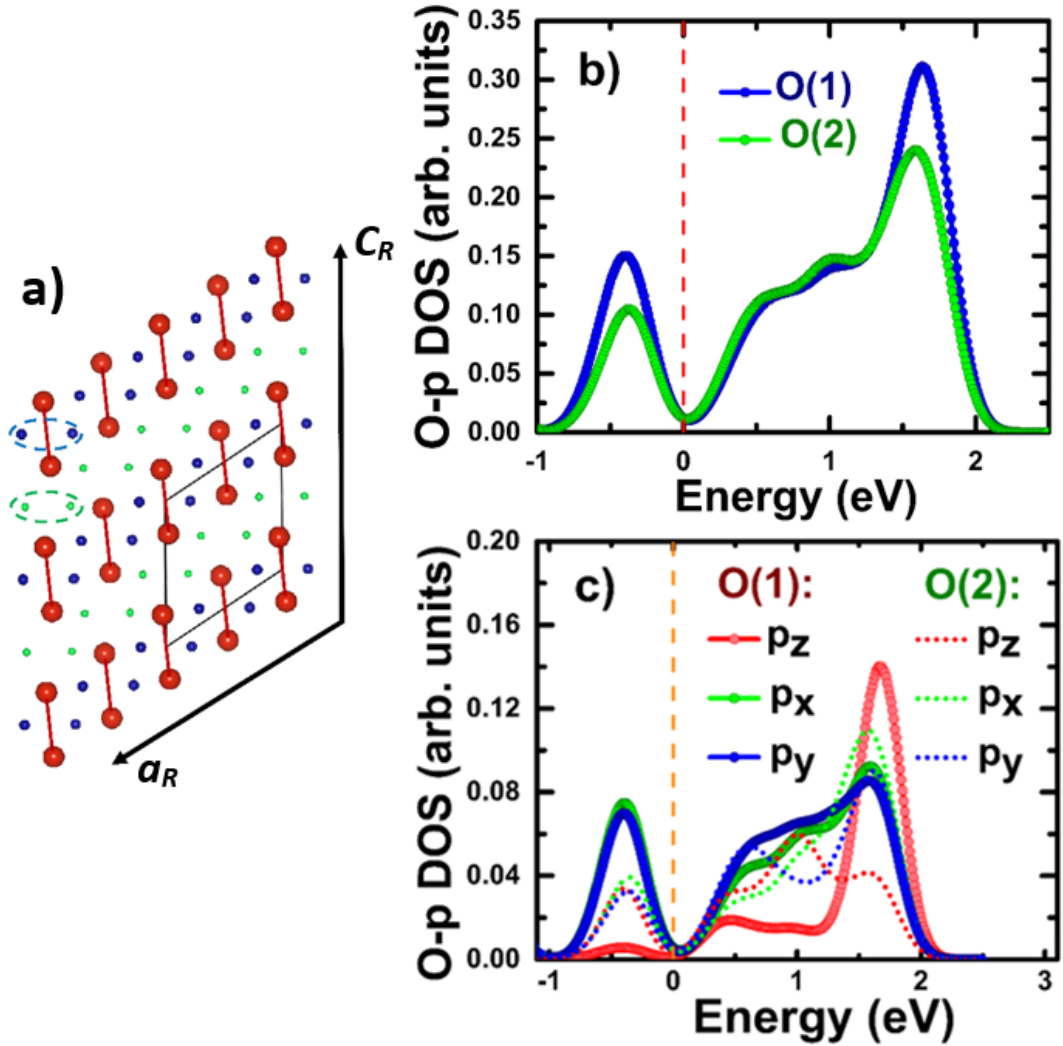

**Figure 3.** (a): Dimerized vanadium atom system (red balls and sticks) and two nonequivalent oxygen atoms (blue – close to a vanadium dimer, green – close to a non-dimerized vanadium-atom pair). Representative non-equivalent oxygen atoms are encircled by dashed ovals. (b): total p-electron oxygen-atom DOS for two nonequivalent atoms shown in Fig. (a). (c): projected DOS that corresponds to the total DOS shown in Fig. (b).

Indeed, as it follows from our calculations for the static (Coulomb) electric potential induced by the pulse leads to a charge redistribution on the vanadium and oxygen chains, Figs. 4 and 5 (details of the calculations are given in SI, Section E).

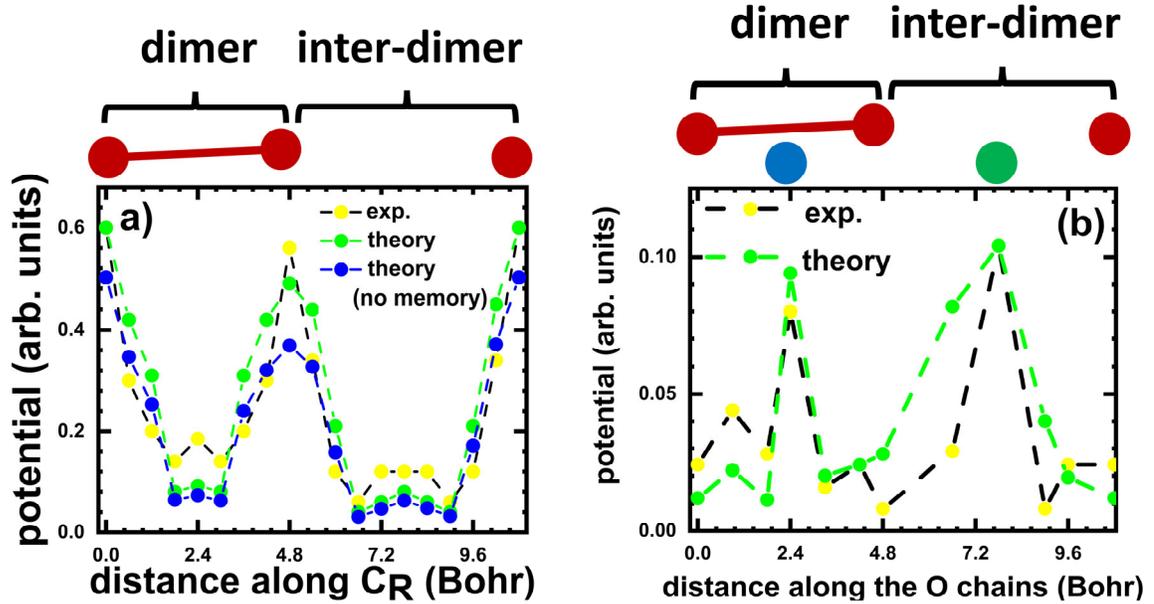

**Figure 4.** Pulse-induced modification of the static potential along the vanadium chain atoms line (a) and in the same direction but along the oxygen atoms line (b). Both theoretical and experimental data [22] are shown in ovals. The lines are approximate fitting of the results. The pulse parameters are given in the caption to Fig.2.

Most important, there is a rather strong modification of the potential (and hence charge density, Fig. 5) on both types of the oxygen atoms, with stronger modification on atoms close to the dimerized vanadium pairs. Surprisingly, there is also a large change of the potential (charge density) after the excitation in the inter-dimer region. This results suggest the idea of vanadium-chain-type of metallization (and hence, anisotropic conductivity) in agreement with data shown in Fig. 2. Our calculations also show that memory effects play a rather important role in the charge redistribution, and hence in modification of the potential along the chains of vanadium atoms. Namely, the values of the theoretical static potential obtained with the XC kernel with memory are approximately 2.5 closer to the experimental data, as compared adiabatic kernel results (Fig. 4a).

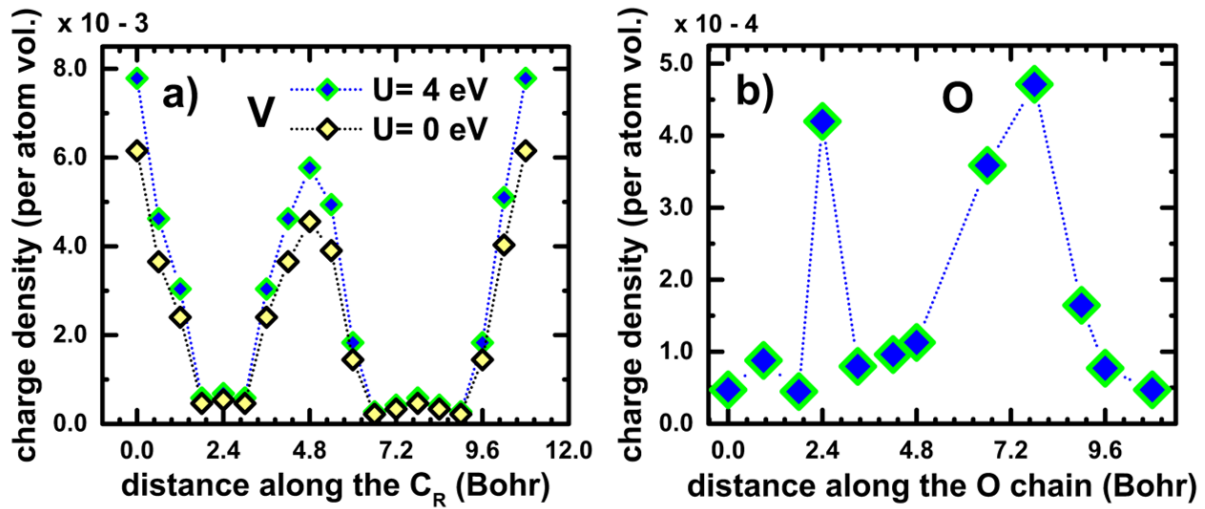

**Figure 5.** The charge density redistributions corresponding to the potentials change in Fig. 4.

To get a deeper insight on the mechanism of the non-homogeneous charge redistribution on the oxygen atoms after excitation, we performed a TDDFT analysis of the charge dynamics on different oxygens. As our results show, in agreement with Fig. 5 the inter-dimer atoms accommodate more excited charge (Fig. 6a). It must be noted that correlation effects do not play a decisive role at the fs timescale on these atoms, thus the charge dynamics is mostly defined by the pulse shape (the charge density shows a decrease as pulse decays; we have used the 5 fs pulse for a better visualization of the difference of the charge densities). Orbital-resolved analysis shows that the difference in the charge density for two atoms is mostly defined by the difference in the occupancy of the $p_z$ orbitals (Fig. 6b). Indeed, closer look at the projected DOS for two oxygen atoms (Fig. 3) shows that the $p_z$-DOS for the inter-dimer atom is much larger near the bottom of the conduction band, favoring the charge transfer. Moreover, the $p_z$-orbital "dumbbells" of the inter-dimer oxygens face the "free" (un-dimerized) sides of the vanadium atoms. Thus, contrary to the intra-dimer atoms, the inter-dimer oxygens are in a better position to accommodate the dangling (unbound) charges of the vanadium atoms. These facts lead to an unusual charge ordering in oxygen chains, in agreement with experiment.[23]

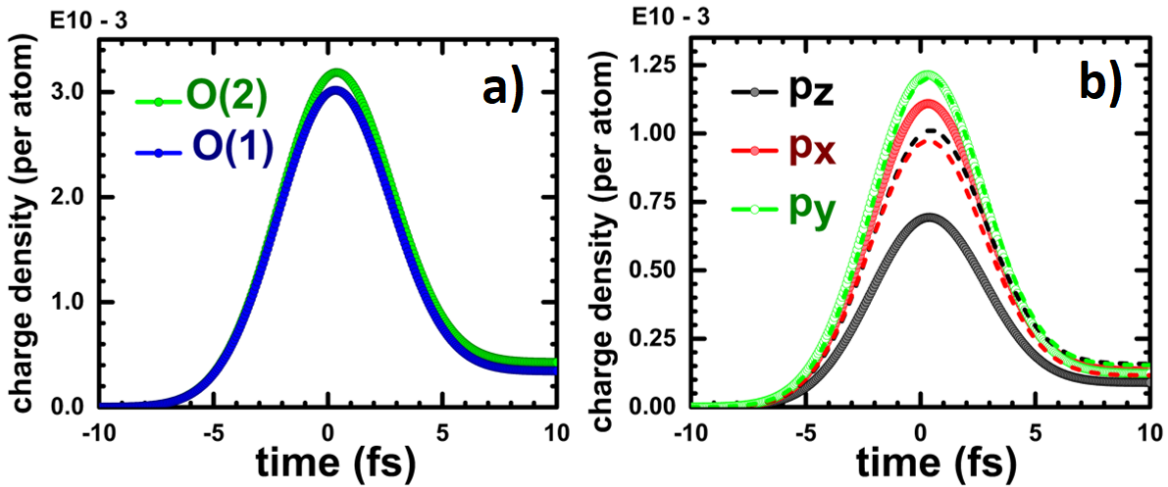

**Figure 6.** (a) Pulse-induced modification of the charge density on the in-dimer (1) and intra-dimer (2) vanadium atoms. The pulse duration is 5fs and the field strength E=1eV/A. (b) The same as in (a) for projected p-orbital DOS for the oxygen in-dimer (solid) and inter-dimer (dashed) atoms.

*Conclusions.* – In this work, we have applied the nonadiabatic TDDFT+DMFT approach to study the ultrafast electronic response in the $M_1$ insulating phase of $VO_2$. It was shown that the initial metallization preferably takes place along the vanadium atom dimer chains. Analysis of the time-dependence of the size of excited metallic domains shows that the domains grow as a power function of time, that results in a power-like time-dependence of the fs conductivity – a result that can be tested experimentally. It was also established that nonequivalence of the "dimer" and "inter-dimer" oxygen atoms lead to a nontrivial modulation of the excited charge density on chains of these atoms. The memory effects play an important role in the nonhomogeneous ultrafast charge dynamics in $VO_2$. Our calculations show that the ultrafast charge transfer between the vanadium and oxygen states takes place at the femtosecond timescale, suggesting that the experimental conclusions that the IMT in VO$_2$ may happen within tens of femtoseconds might be correct.

We plan to extend the analysis on the case of longer times, including the phonon and other temperature effects (like transformations of the lattice structure).

---

*Acknowledgements*—We thank R.D. Averitt, W.H. Brito and J. Staehler for very helpful communications. We acknowledge DOE for a partial support under grant DOE-DE-FG02-07ER46354. J.M.G.H. would like to thank CONACYT (Mexico) for a support through the Postdoctoral Fellowship Program (Scholarship # 23210).

# SUPPLEMENTARY INFORMATION

# Spatially non-homogeneous metallization of VO$_2$:

# a TDDFT+DMFT analysis

Jose Mario Galicia-Hernandez,[a,b] Volodymyr Turkowski,[a] Gregorio Hernandez-Cocoletzi[b] and Talat S. Rahman[a,*]

[a]Department of Physics, University of Central Florida, Orlando, FL 32816, USA
[b]Instituto de Fisica Ing. Luis Rivera Terrazas, Benemerita Universidad Autonoma de Puebla, Puebla 72550, Mexico,

## A. Details of the DFT+DMFT calculations

The *DFT+DMFT* calculations were performed by solving an effective five d-band Hubbard problem with the following Hamiltonian:

$$H = -\sum_{i,j,l,m,\sigma} t_{ij,\sigma}^{lm} c_{i\sigma}^{l+} c_{j\sigma}^{m} - \mu \sum_{i,l,\sigma} n_{i\sigma}^{l} + U \sum_{i,l} n_{i\uparrow}^{l} n_{i\downarrow}^{l} + (U-J) \sum_{i,l\neq m} n_{i\uparrow}^{l} n_{i\downarrow}^{m} + (U-2J) \sum_{i,l\neq m,\sigma} n_{i\sigma}^{l} n_{i\sigma}^{m}$$

$$-V \sum_{i,l,\sigma} v_{i,i+d}^{l} c_{i\sigma}^{l+} c_{i+d\sigma}^{l}, \qquad (A.1)$$

where $c_{i\sigma}^{l+}$ ($c_j^m$) and $n_{i\sigma}^m$ are the electron creation (annihilation) and the number operators, $t_{ij,\sigma}^{lm}$ are the hopping parameters (i,j are the site- and l, m are the orbital indices; $\sigma$ is the spin index), which can be obtained from the DFT spectrum; $\mu$ is the chemical potential. In the third-to-fifth terms, U=5eV, U-J=4eV and U-2J=3eV are the on-site Coulomb repulsion parameters. They correspond to the following interactions: U – same-orbital opposite-spin repulsion, U-J different-orbital opposite-spin repulsion, and U-2J - different-orbital same-spin. Finally, the last term is the Fock-approximated part of the Hamiltonian that describes the inter-site same-orbital same-spin interaction within the vanadium dimers (i and i+d are the intra-dimer atom position indices). In this term, V=1eV is the corresponding repulsion energy and $v_{i,i+d,\sigma}^l = \langle c_{i\sigma}^{l+} c_{i+d\sigma}^l \rangle$ is the occupation matrix. It is assumed that the Hartree part of this energy is included into the DFT results and other intra-dimer interactions are small (see, e.g., Ref. [1]). The choice of the values of U, J and V is motivated by the previous experimental and theoretical results.

The DMFT electronic spectrum and DOS were obtained by calculating the spin and orbital dependent single-particle Green's function

$$G_{i\sigma j\sigma'}^{ll'}(r,t;r',t') = -\langle \hat{T} c_{i\sigma}^{l}(t) c_{j\sigma'}^{l'+}(t') \rangle, \qquad (A.2)$$

where $\hat{T}$ is the time-ordering operator. In the DMFT case this function has the following form in the momentum-frequency representation:

$$G_{ab}(k,\omega) = \left[\omega - \hat{\varepsilon}_k - \hat{\Sigma}(\omega)\right]_{ab}^{-1}, \qquad (A.3)$$

where a and b are the combined (orbital-spin) quantum indices, $\hat{\varepsilon}_k$ is the free spectrum (diagonal) matrix. The self-energy $\hat{\Sigma}(\omega)$ in our case of VO$_2$ was taken to be a momentum-independent matrix with two atomic (vanadium dimer) site indices and five orbital indices (with the matrix elements corresponding to two atoms within the vanadium dimer - the cluster DMFT approximation[2]). Momentum-independence of the self-energy matrix is equivalent to neglecting the spatial fluctuations beyond the vanadium dimers and is a great advantage of the DMFT approximation, which allows one to solve the problem in principle exactly or by using well-controlled approximations. Namely, in the DMFT the many-atom and many-electron problem is reduced to the "impurity problem" – a problem of a single site (or a cluster of sites) in presence of an effective bath of electrons. To solve the impurity problem, we use the Multi-Orbital Iterated-Perturbation Theory (MO-IPT) approximation,[2,3] where the impurity electron self-energy depends on the interaction parameters U, J and V and on the effective bath field, and is calculated by employing a generalized perturbation theory in terms of U, J and V (for details of the DMFT calculations see, e.g., Ref. [2]). The DMFT-corrected electron spectrum was obtained from the Green's function Eq. (A.3) by solving the "pole equation" $\omega - \hat{\varepsilon}_k - \hat{\Sigma}(\omega) = 0$, and the corresponding DOS was calculated as $A(\omega) = -\frac{1}{\pi}\sum_a \int dk \, \text{Im} \, G_{aa}(k,\omega)$.

Since for the TDDFT (next) part, we need the XC kernel obtained from the DMFT charge local-in-space susceptibility, we present brief details of the calculation of this function.

Namely, the charge susceptibility is defined as

$$\chi(r,t;r',t') = -\langle \hat{T}n(r,t)n(r',t')\rangle = \sum_{l,l',\sigma,\sigma'} \chi_{\sigma\sigma'}^{ll'}(r,t;r',t'), \qquad (A.4)$$

and as it follows from the last equality, it can be obtained by calculating the orbital-spin susceptibilities:

$$\chi_{\sigma\sigma'}^{ll'}(r,t;r',t') = -\langle \hat{T}n_\sigma^l(r,t)n_{\sigma'}^{l'}(r',t')\rangle. \qquad (A.5)$$

Once the last quantities, calculated in the local-in-space ($r = r'$) one-loop approximation and Matsubara frequency representation as

$$\chi_{ab}^{(0)}(\omega_n) = -T\sum_m G_{ab}(\omega_n + \omega_m)G_{ba}(\omega_m), \qquad (A.6)$$

are found one obtains the total local-in-space one-loop susceptibility $\chi^{(0)}$ by substituting the time-transformed Eq.(A.6) the last in Eq.(A.4) (we have used here and in some equations below the Matsubara frequency representation, where for fermions $\omega_n = \pi T(2n+1)$, T is the temperature and n are integer numbers). Then, in order to get the vertex correction to the on-loop susceptibility, we use a computationally inexpensive yet rather powerful static ladder approximation,[4,5] that gives

$$\chi(\omega_n) = \frac{\chi^{(0)}(\omega_n)}{1+U_{ch}\chi^{(0)}(\omega_n)}, \qquad (A.7)$$

where $U_{ch}$ is an effective parameter, that correspond to the static ladder approximation for the vertex function,

$$\Gamma(\omega_m, \omega_l, \omega_n) \cong U_{ch}. \qquad (A.8)$$

The value of $U_{ch}$ is fixed by requiring that the total susceptibility (A.7) to satisfy the sum rule

$$T\sum_m \chi(\omega_n) = n(1-n) + 2D, \qquad (A.9)$$

where n is the on-site electron density and

$$D = \langle n_{i\uparrow} n_{i\downarrow} \rangle = \left(\frac{n}{2}\right)^2 + T \sum_{a,n} \Sigma_{aa}(\omega_n) G_{aa}(\omega_n) \qquad (A.10)$$

is the one-site electron double occupancy.

Approximation (A.7) is exact in the linear in U approximation (when $U_{ch} = U$). It is also capable to reproduce many important features of the exact DMFT solution at large U's, including superconductivity in the two-dimensional Hubbard model (for details and references, see, e.g., Ref.[5]).

**B. Details of the TDDFT calculations.**

To study the response of the system, we solved the TDDFT Kohn-Sham equation:

$$i\frac{\partial \Psi(r,t)}{\partial t} = \left[-\frac{\nabla^2}{2m} + V_{ion}(r) + V_H[n](r,t) + V_{XC}[n](r,t) + V_{exc}(r,t)\right] \Psi(r,t), \qquad (B.1)$$

where $V_{ion}(r)$, $V_H[n](r,t)$, $V_{exc}(r,t) = -e\vec{E}(t) \cdot \vec{r}$ are the ion, Hartree and external ($\vec{E}(t)$ is the pulse field) potentials, respectively. $V_{XC}[n](r,t)$ is the XC potential, which we use the linear in the fluctuations of the charge density approximation:

$$V_{XC}[n](r,t) \approx V_{XC\sigma}[n](r, t=-\infty) + \int_{-\infty}^{t} dt' \int dr' f_{XC}(r,t;r't') \delta n(r',t'). \qquad (B.2)$$

where $V_{XC\sigma}[n](r, t=-\infty)$ is the static (unperturbed) part of the XC potential and

$$f_{XC}(r,r';t,t') = \frac{\delta V_{XC\sigma}[n](r,t)}{\delta n(r',t')} \qquad (B.3)$$

is the XC kernel. The $V_{XC\sigma}[n](r, t=-\infty)$ contributes only to the "correlation" correction of the DFT spectrum (e.g., opening the gap), i.e. we assume that $\left[-\frac{\nabla^2}{2m} + V_{ion}(r) + V_H[n](r, t=-\infty) + V_{XC}[n](r, t=-\infty)\right] \psi_{\sigma k}^l(r) = \varepsilon_{\sigma k}^l(r) \psi_{\sigma k}^l(r)$. Thus, we don't need its explicit form in the density-matrix TDDFT calculations.

The DMFT XC is obtained from the non-interacting $\chi^{(0)}$ ($U_{ch}$) and interacting susceptibilities $\chi$, Eq. (A.7), as

$$f_{XC}(r, r', \omega) = \delta(r - r')\left[\chi^{(0)-1}(\omega) - \chi^{-1}(\omega)\right] - \frac{1}{|r - r'|} \qquad (B.4)$$

(in the real frequency representation).

Since the XC potential depends on the fluctuation of the electron density:

$$\delta n(r,t) = \sum_{l,\sigma} [|\psi_\sigma^l(r,t)|^2 - |\psi_\sigma^l(r, t=-\infty)|^2], \qquad (B.5)$$

Eqs. (B.1),(B.5) must be solved self-consistently.

Two comments have to be made regarding the approximation above. The linear response approximation for the XC potential (B.2) is valid when the fluctuating charge density is small comparing to the total number of electrons, which is valid in many cases even with a strong perturbation. This approximation is capable to describe multi-electron effects, in particular due to the frequency dependence of the XC kernel, that takes the memory effects into account. The second approximation – spatial locality of the DMFT part of the XC kernel (B.4), corresponds to the case when only the local (on-site) dynamical interactions of the charges are taken into account, which may be regarded as a reasonable choice for localized d- and f-orbitals, which as DMFT studies show is in most cases a sufficient approximation. To take into account spatial fluctuations more accurately one can include the non-local space- (momentum-) dependence of the kernel through the momentum dependence of the vertex function (A.8).[6]

To solve the TDDFT equation (11), we used the density-matrix formalism, where the time-dependent wave function is expanded in terms of the relevant static d-orbital KS wave functions:

$$\Psi_{\sigma k}(r, t) = \sum_{l,\sigma} c_{\sigma k}^l(t) \psi_{\sigma k}^{l(0)}(r) \tag{B.6}$$

(l and $\sigma$ are the orbital and spin indices). In this case, the problem reduces to solving the Liouville equation for the time-dependent density-matrix elements constructed from the time-dependent c-coefficients:

$$\rho_{\sigma k}^{lm}(t) = c_{\sigma k}^l(t) c_{\sigma k}^{m*}(t) \tag{B.7}$$

(see, e.g., Ref. [49]). Namely, the elements of the density matrix satisfy the equation

$$i \frac{\partial}{\partial t} \rho_k^{lm}(t) = [H(t), \rho(t)]_k^{lm}, \tag{B.8}$$

where

$$H_k^{lm}(t) = \int d^3r \, \psi_k^{l(0)*}(r) H(r,t) \psi_k^{m(0)}(r) \tag{B.9}$$

are the matrix elements of the KS Hamiltonian with respect to the basis wave functions.

The excited charge density is equal to the sum of the all conduction-band diagonal matrix elements (B.7) over momentum.

**C. Growth of the metallic domains**

To model spatially-nonhomogeneous metallization, and in particular to test the scenario above, we consider another type of excitation – a homogeneous-in-space electric field turned on at time t=0, $\vec{E}(t) = \vec{E}_0 \theta(t)$. The presence of a constant field makes the insulating system unstable towards fluctuations, that may lead to metallic domain nucleation and spinodal decomposition (separation of the metallic and insulating areas) at the femtosecond timescale. Due to complexity of the problem, we apply combined analytical (scaling analysis[7,8]) and (as the next step) numerical approaches to consider the ultrafast response in this case.

As the analytical part, we transform the KS equation to the equation for the time-dependent elements $c_{\sigma k}^c(t)$ the KS wave function, Eq. (B.1). These elements define the excited charge density for each "conduction" d-band (more precisely, the density is the square of the modulus of these elements). To derive this equation, one has to substitute the wave function expansion (B.6) into the KS equation (B.1), multiply by a complex-conjugated static KS wave function, e.g. $\psi_{\sigma q}^{m(0)*}(r)$, and perform the spatial integration. Then, after simple transformations and simplifications (see the text below the next equation), one can obtain the following equation:

$$i\frac{\partial c_k^c(t)}{\partial t} = \left[\varepsilon_k^c + \sum_{c',q} F_{kq}^{cc'} \int_{-\infty}^t f_{XC}(t-t')|c_q^{c'}(t')|^2 dt'\right] c_k^c(t) + \vec{d}_k^c \vec{E}(t), \tag{C.1}$$

where $\vec{d}_k^c = \sum_v \vec{d}_k^{cv}$ is total dipole moment for given band c obtained after summation over all (five) valence bands, and $F_{kq}^{cc'} = \int dr |\psi_k^{c(0)}(r)|^2 |\psi_{\sigma q}^{c'(0)}(r)|^2$. The only nontrivial approximation made in the exact equation for $c_k^c(t)$ was approximation $F_{kq}^{cc'}$ for four-state function $\int dr \psi_{\sigma k}^{a(0)*}(r) \psi_{\sigma q}^{b(0)}(r) \psi_{\sigma p}^{c(0)*}(r) \psi_{\sigma Q}^{d(0)}(r) \approx \delta^{ab}\delta_{kq}\delta^{cd}\delta_{pQ} \int dr |\psi_k^{a(0)}(r)|^2 |\psi_{\sigma q}^{c(0)}(r)|^2$, i.e. choosing the largest matrix elements.

Equation (C.1) is still difficult to solve exactly, thus we make further simplification by ignoring momentum dependence in functions $F_{kq}^{cc'}$, putting $F_{kq}^{cc'} \approx F_{00}^{cc'}$ (the lowest-order $\vec{k}\cdot\vec{p}$ approximation) and introducing an average interaction $\bar{F}$, defined as $\sum_{c'} F_{00}^{cc'} \sum_q |c_q^{c'}(t')|^2 = \bar{F}\sum_q|c_q^{c'}(t')|^2$. Averaging the functions $F_{00}^{cc'}$ over the conduction orbitals gives $\bar{F} = 0.23$ (in atomic units). Then, equation (D.1) transforms to:

$$i\frac{\partial c_k^c(t)}{\partial t} = \left[\varepsilon_k^c + \bar{F} \int_0^t f_{XC}(t-t') \sum_{c',q}|c_q^{c'}(t')|^2 dt'\right] c_k^c(t) + \vec{d}_k^c \vec{E}(t). \tag{C.2}$$

This equation can be regarded as a generalized form of the $\phi^4$-theory equation $i\frac{\partial \phi_k(t)}{\partial t} = \left[m^2(t) + \frac{\lambda}{6}\phi_k^2(t)\right]\phi_k(t) + E(t)$ (see, e.g., Ref.[8]). There are, however, very important differences between the two theories: our fields are complex; equation (D.2) is the first order in time equation (contrary to the second order one in the $\phi^4$-theory); instead of the mass m in the linear (at zero momentum) term $m^2\phi_k$ (for perturbation, the sign of $m^2$ is flipped to generate an instability) it includes positive constant gap $E_g^c$, (incorporated into $\varepsilon_k^c$). Finally, the interaction term contains memory effects, contrary to the instant $\frac{\lambda}{6}\phi_k^3(t)$ term in the $\phi^4$-theory.

In our theory, the instability of the ground state is caused by modification of the interaction term, not by flipping of the sign in front of the "mass" term $\varepsilon_k^c c_k^c(t)$, after the electric field is turned on. This can be easily seen from the following qualitative analysis. The potential part of the equation $E_g^c + \bar{F}\int_0^t f_{XC}(t-t')\sum_{c',q}|c_q^{c'}(t')|^2 dt'$, can be approximated by $m^2(t) + \bar{F}\int_0^t f_{XC}(t-t')\delta n(t')dt'$, where $m^2(t) = E_g^c + \bar{F}\bar{n}\int_0^t f_{XC}(t-t')t'$, $\bar{n}$ is an average in time value of the excited density $\sum_{c',q}|c_q^{c'}(t')|^2$ and $\delta n(t') = \sum_{c',q}|c_q^{c'}(t')|^2 - \bar{n}$ is the charge flcutuation. Due to oscillating structure of the function $f_{XC}(t-t')$, which is negative at large intervals of time, one can see that $m^2(t)$ may flip the sign as soon as the excited charge density $\bar{n}$ is not equal zero. One can regard such instability as dynamically-generated. Thus, one may expect creation and expansion of the metallic domains.

Solving Eq.(C.2) for $c_q^l(t)$, transforming the result to the real space representation $c^l(\vec{r},t)$ (we used the real-space domain of the size of $3\times 3\times 3$ of the original unit cell (Fig. 1a)), one can calculate the correlation function for the sum of different orbital contributions $c(\vec{r},t) = \sum_l c^l(\vec{r},t)$, $S(\vec{r},t) = \langle c(\vec{r},t)c(\vec{r},0)\rangle$. Then by solving equation $S(\vec{R},t) = $ constant, one can find the time dependence of the radius (size) of the excited domain $\vec{R}(t)$.

## Appendix D. Calculations of the conductivity

To calculate the conductivity as function of time we use our results for the time-dependence of metallic domains and Bruggeman's formula [9] which connects these two quantities:

$$V_m(t) \frac{\sigma_{met} - \sigma(t)}{\sigma_{met} + 2\sigma(t)} + \left(1 - V_m(t)\right) \frac{\sigma_{ins} - \sigma(t)}{\sigma_{ins} + 2\sigma(t)} = 0, \quad (D.1)$$

where $V_m(t)$ is metallic fraction of the system (defined by the domain radii in different directions), and $\sigma_{ins}$, $\sigma_{met}$ and $\sigma(t)$ are the insulating metallic, and time-dependent conductivities. We used $\sigma_{ins} = 0$, which gives

$$\sigma(t) = \sigma_{met} \left(\frac{3}{2} V_m(t) - \frac{1}{2}\right). \quad (D.2)$$

It was also assumed that the density of metallization centers is $0.5 nm^3$.

## E. Calculation of the growth of the metallic domains

To calculate the spatial distribution of the excited charge density along the vanadium and oxygen atom lines, we have used the values of the momentum-dependent c-coefficients at final moment of time $t_f = 10fs$ (see Eqs. (B.6),(B.7) and Fig.6 for the corresponding charge dynamics), substituted these coefficients into Eq.(B.6) to calculate the corresponding space dependent wave function and after that – the excited charge density as difference between the final and initial charge densities

$$\delta n(\vec{r}) = \sum_{k,\sigma} \left(|\Psi_{\sigma k}(\vec{r}, t_f)|^2 - |\Psi_{\sigma k}(\vec{r}, t_0)|^2\right). \quad (E.1)$$

Using the results for the excited charge density, we plotted this function along the V and O lines in z-direction.

The modification of the static electric potential along these chains was calculates as

$$\delta V(\vec{r}) = \int \frac{\delta n(\vec{r}')}{|\vec{r} - \vec{r}'|} d\vec{r}', \quad (E.2)$$

where the integration was performed only over the unit cell, assuming that the potential modification mostly comes from the atoms near to the chain (due to the same reason, the screening effects were neglected).